\documentstyle[aps,prb,multicol]{revtex}

\begin{document}
\input{epsf}
\draft \preprint{}
\title{Transport signatures of correlated disorder in a
two-dimensional electron gas}
\author{T. Heinzel\cite{byline1}, R. J\"aggi, E. Ribeiro\cite{byline2},
M.v. Waldkirch, and K. Ensslin}
\address{Solid State Physics Laboratory, ETH Z\"{u}rich, 8093
Z\"{u}rich,  Switzerland\\}
\author{S. E. Ulloa}
\address{Department of Physics and Astronomy, Ohio University, Athens,
OH 45701-2979, USA\\}
\author{G. Medeiros-Ribeiro\cite{byline2}, and P. M. Petroff}
\address{Materials Department, University of California, Santa
Barbara, CA 93106, USA\\}
\date{\today}
\maketitle
\begin{abstract}
We report electronic transport measurements on two-dimensional
electron gases in a Ga[Al]As heterostructure with an embedded
layer of InAs self-assembled quantum dots. At high InAs dot
densities, pronounced Altshuler-Aronov-Spivak magnetoresistance
oscillations are observed, which indicate short-range ordering of
the potential landscape formed by the charged dots and the strain
fields. The presence of these oscillations coincides with the
observation of a metal-insulator transition, and a maximum in the
electron mobility as a function of the electron density. Within a
model based on correlated disorder, we establish a relation
between these effects.
\end{abstract}

\begin{multicols} {2}
The optical and electronic properties of self-assembled quantum
dots (SAQDs) have been widely studied in the past few
years.\cite{Leonard1993,Moison1994,Marzin1994,Drexler1994} These
experiments have been performed on individual SAQDs or on
two-dimensional dot arrays. Recently, however, substantial
progress in fabrication technology has led to self-ordered growth
of such quantum dots in two
\cite{Xie1995,Tersoff1996,Romanov1999,Pinczolitis1999,Lee2000} and
even three \cite{Springholz1998} dimensions. The ordering has been
detected by scanning force microscopy as well as by X-ray
diffraction. These studies suggest that designing electronic
minibands by self-ordered growth of SAQDs may be feasible.

Here, we report experimental evidence in \emph{electronic
transport}, that InAs SAQDs embedded at the heterointerface of a
Ga[Al]As heterostructure, \cite{Ribeiro1998} are capable of
creating a potential landscape with substantial spatial and
amplitude correlations. In our samples, the SAQDs have been grown
on an atomically flat GaAs [100] surface. The charged SAQDs,
combined with the induced strain fields, act as repulsive
scatterers in a two-dimensional electron gas (2DEG). At
sufficiently high dot densities, several effects are observed to
occur in concert: (1) Magnetoresistivity measurements show
pronounced, periodic oscillations, which are interpreted as
Altshuler-Aronov-Spivak oscillations. In addition, under certain
parametric conditions, a resistivity minimum around zero magnetic
field is observed, superimposed with the weak localization peak.
(2) We observe a maximum in the electron mobility as a function of
the electron density. (3) The temperature dependence of
$\rho_{xx}$ as a function of electron density shows a transition
from metallic behavior to insulating behavior,\cite{Ribeiro1999}
an effect that has been interpreted as a signature of a
metal-insulator transition in two dimensions. We emphasize that in
the samples under study, these transport features always occur
together.  As we describe below, this complex experimental
behavior is a signature of the potential correlations
\cite{Buks1994} in the 2DEG
provided by the SAQD potential landscape.\\
The samples have been grown by molecular beam epitaxy. A layer of
InAs SAQDs has been embedded in the GaAs buffer layer 3 nm below a
$GaAs - Al_{0.3}Ga_{0.7}As$ - heterointerface, at which the 2DEG
resides. The density $n_{QD}$ of the InAs SAQDs varies across the
wafer. $n_{QD}$ has been determined, within a factor of two, by
transmission electron microscope (TEM) studies, as well as by
low-temperature transport experiments. For details of the sample
growth and characterization, see Ref.\ \onlinecite{Ribeiro1998}.
The wafer was cleaved into pieces, and conventional Hall bar
geometries were patterned by optical lithography. Voltages applied
to a homogeneous top gate electrode were used to tune the electron
density $n_{e}$ of the 2DEG. Transport experiments were carried
out in a $^{3}$He/$^{4}$He dilution refrigerator and in a standard
$^{4}$He cryostat at temperatures between 60 mK and 20 K. The
resistivity was measured in a standard 4-terminal setup, probing
sample areas of $40\mu m \times 100\mu m$. For $n_{QD}\lesssim
2\times 10^{14}m^{-2}$, the electron gases behave like ordinary
low-mobility samples, in which the mobility is dominated by the
SAQDs.\cite{Ribeiro1998}\\
At higher dot densities of $n_{QD}\approx 2.5\times 10^{14}m^{-2}$
(i.e. the samples labelled by 14a, 14b and 24b, see inset in Fig.\
1a), we find pronounced oscillations as a function of a magnetic
field $B$ applied perpendicular to the plane of the 2DEG. The
oscillations are periodic in $B$, with a period of $\delta B =
400\pm 20$ mT in sample 14a, which increases to $560\pm 20$ mT in
sample 24b. The period is independent of $n_{e}$ (Fig.\ 1a). The
relative oscillation amplitude $\delta\rho /\rho$ increases as
$n_{e}$ is increased, with a maximum of $\delta\rho /\rho \approx
0.12$. Measurements in tilted magnetic fields show that the
oscillations are determined solely by the magnetic field component
perpendicular to the plane of the 2DEG, and have therefore a
purely dynamical (no spin) origin. These oscillations are
superimposed on a broad resistivity background, which vanishes
around $B=3T$. We do not observe Shubnikov-de Haas oscillations
for filling factors larger than 4.
\begin{figure}[t]
\centerline{\epsfxsize=3.0in \epsfbox{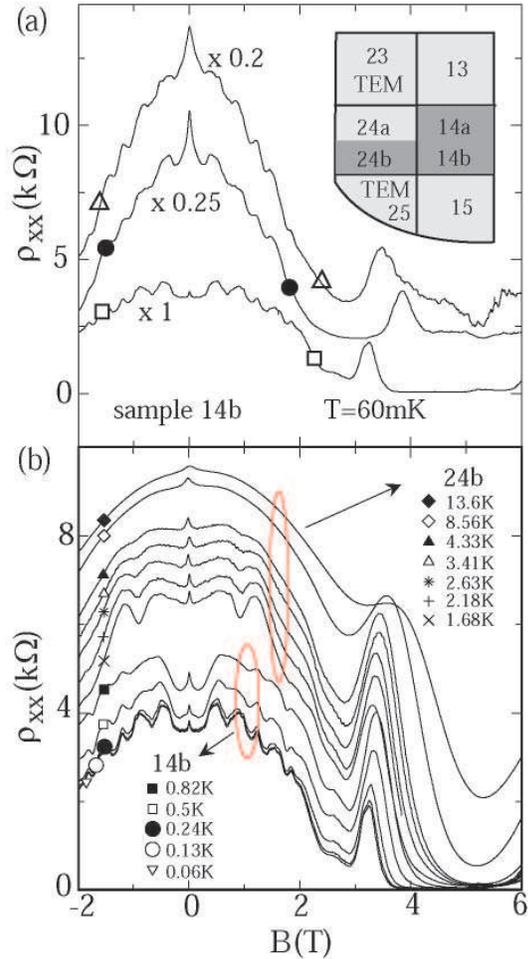}} \caption {(a)
Magnetoresistivity of sample 14b for $n_{e}=2.37\times
10^{15}m^{-2}(\Box)$, $1.16\times 10^{15}m^{-2}(\bullet)$, and
$0.97\times 10^{15}m^{-2}(\triangle)$; the resistivity data for
the two lower electron densities are scaled as indicated.
Pronounced magneto-oscillations are observed for $|B|<2T$. Their
period and phase is independent of $n_{e}$. In addition, a
positive magnetoresistivity is observed for $|B|<0.5T$, with the
weak localization peak superimposed. The inset shows a sketch of
the wafer under study. $n_{QD}$ increases from top to bottom (edge
of the wafer), and has been estimated by TEM studies as $n_{QD}=
2.5\times 10^{14}m^{-2}$ in sample 23, and $n_{QD}= 3.0\times
10^{14}m^{-2}$ in sample 25, respectively. The
magneto-oscillations are observed on samples shown in dark gray.
(b) Temperature dependence of the magnetoresistivity in samples
14b and 24b.} \label{CFFig1}
\end{figure}

In addition to a weak localization peak, we observe for large
electron densities a well-pronounced minimum in the
magnetoresistivity around $B=0$, which extends up to a few hundred
mT. In contrast to the magneto-oscillations, this minimum shows no
significant temperature dependence up to about 2.5 K (Fig.\ 1b).
Figure 2, shows additional observations that coincide with the
presence of the features discussed above: (i) a fixed point is
observed in the temperature dependence of $\rho_{xx}(n_{e})$ at
$B=0$, separating a region of high electron density
$(n_{e}>1.2\times10^{15}m^{-2})$ with a metallic temperature
dependence, from a low density region showing an insulating
temperature dependence.\cite{Ribeiro1999} (ii) the electron
mobility $\mu$ shows a maximum as a function of $n_{e}$ (upper
inset in Fig. 2).
\begin{figure}
\centerline{\epsfxsize=3.0in \epsfbox{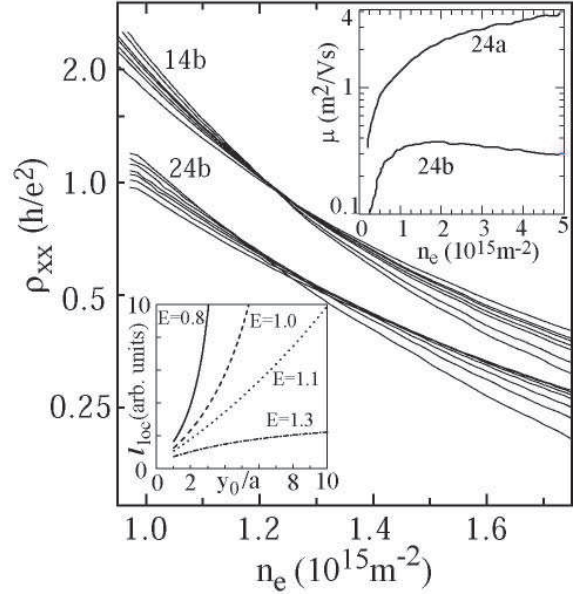}}
 \caption {Metal-insulator transition as observed as a function of $n_{e}$
and temperature in two samples. The upper inset shows the
electronic mobility $\mu$ as a function of $n_{e}$ for two
samples. Sample 24b shows a maximum in the electron mobility, in
contrast to sample 24a. Lower inset: localization length vs
correlation length $y_{0}$ for the correlation function discussed
in text in a 1D system, and selected energy values. $l_{loc}$
increases rapidly for larger $y_{0}$, with a rate that depends on
the energy $E$ of the state. An even faster enhancement of
$l_{loc}$ is anticipated for a 2D system. Algorithm to calculate
$l_{loc}$ is described in Ref.\ 19.} \label{CFFig2}
\end{figure}
 The presence of these effects depends on the
cooldown cycle. Hence, the arrangement of charges in the potential
landscape during cooldown is a crucial factor. However, we have
been unable to establish a systematic dependence on the cooldown
speed or on the biasing conditions during cooldown. The
simultaneous occurrence of the transport features described here
indicates a possible unifying explanation which is developed
below.\\
In the following, we focus on sample 14a and interpret its
behavior, starting with the periodic magneto-oscillations. The
experiments clearly suggest a magnetic flux effect as its origin.
It is well-known that Aharonov-Bohm oscillations with period
$\Delta B =h/eA$ ($A$ denotes the enclosed area) ensemble-average
to zero, since the phase of each interference loop at $B=0$ is
loop-specific.\cite{Umbach1986} In contrast,
Altshuler-Aronov-Spivak (AAS) oscillations \cite{Spivak1981}
always show zero phase shift at $B=0$, and consequently survive
ensemble-averaging. We thus conclude that the observed
oscillations are AAS oscillations with a characteristic enclosed
area of $5.2 \times 10^{-15}m^{2}$.\\
Similar oscillations have been recently detected in hexagonal
antidot lattices, while no signature of such oscillations has been
found in rectangular lattices with otherwise comparable
parameters.\cite{Nihey1995} In Ref. \onlinecite{Nihey1995}, the
authors explain this observation in terms of semiclassical
trajectories of the electrons around the antidots: in the
hexagonal array, the probability for the electrons to be scattered
in loops around the antidots is greatly enhanced, leading to a
significant AAS oscillation amplitude,
in contrast to the more open geometry of rectangular lattices.\\

Let us, based on this knowledge and our observations, assume that
the potential landscape generated by the charged SAQDs and the
strain fields reveals at least short range hexagonal ordering to a
significant degree. The measured AAS period would then suggest an
interdot separation of $a\approx 78 nm$, corresponding to a
density of SAQD points of $1.9 \times 10^{14}m^{-2}$. This is in
reasonable agreement with $2\times 10^{14}m^{-2}\lesssim
n_{QD}\lesssim 3\times 10^{14}m^{-2}$, as determined in TEM
studies, although $n_{QD}$ as determined from resistivity studies
is somewhat higher.\cite{Ribeiro1998,Heinzel2000} Further
estimates are consistent with this picture. We have determined the
phase coherence length $\ell_{\phi}$ from fits of the weak
localization peak for $\rho_{xx}<h/e^{2}$ and find
$\ell_{\phi}\approx 500 nm $ at $n_{e}=1.8\times 10^{15}m^{-2}$,
and increasing slightly as $n_{e}$ increases. The electrons are
thus clearly able to eclipse the area of the locally hexagonal
ordered cells in a phase coherent fashion. As a further
consequence, quantum point contacts are formed in between the
lattice points, with a maximum of two occupied 1D spin-degenerate
subbands at the electron densities under investigation. This
explains the quenching of the Shubnikov-de Haas oscillations above
filling factor 4. Using these numbers and assuming that the
hexagonal arrangement contains approximately circular scatterers
of electrostatic radius $r_{QD}$ at the Fermi energy, we can
estimate $r_{QD}\approx 20 nm$. Hence, $2r_{QD}\approx
\frac{a}{2}$, which has been shown to be a particularly effective
geometry for observing AAS oscillations in hexagonal antidot
lattices. \cite{Nihey1995} We must emphasize that this analysis
does not rely on truly long-range order, but rather only on the
local hexagonal relative arrangement of the charged SAQDs. Nor do
we know unambiguously that the ordering is really hexagonal: other
types of correlations in the disorder potential, e.g. possibly a
random trimer superlattice, may give rise to $B$-periodic coherent
backscattering as well. Note, however, that the relative amplitude
of the oscillations is comparable to that one observed in
Ref.\onlinecite{Nihey1995}, which suggests a similar degree of
ordering.\\

We proceed by constructing the relation between the AAS
oscillations, the ``metal-insulator transition" (MIT), and the
maximum in $\mu (n_{e})$. If the density of SAQDs is low, there is
no positional short-ranged order, and even homogeneous charging of
the dots results in a potential landscape dominated by the
disordered dot distribution of charged scattering centers.  As the
interdot separation is still large, the backscattering processes
inherent to the AAS effect are unlikely, and no magneto-resistance
oscillations occur. Similarly, the mobility of the system is
dominated by random scattering events from the dots, yielding an
insulating behavior for all values of electronic density.  As the
density of dots increases, there is an effective repulsive
interdot interaction which facilitates the appearance of
short-ranged order so as to relax the local elastic stress in the
system.\cite{Romanov1999,Lee2001} Upon formation of the 2DEG, the
dots so formed acquire charge with a distribution which is likely
correlated to the loading state of nearby dots (notice that
dot-dot interaction is quite large, as $(2e)^2/\kappa a \approx
6$meV, for doubly-charged dots embedded in GaAs, $\kappa \approx
12$). The short-range order yields enhancement of backscattering
and AAS oscillations, as well as the appearance of relatively
extended states due to the effectively weaker, more correlated,
potential disorder. As the electron density increases, the AAS
oscillations become more prominent, indicating increasing
population of the corresponding states. The resulting electronic
states are then capable of exhibiting a drastically different
transport behavior as a function of electron density: insulating
for low densities, and metallic for higher concentrations.  This
change is the result of the Fermi level sweeping through the
density of states,
sampling ever more extended states.\\
It is worth mentioning that our experiments show a sharp
transition to such a correlated state as a function of $n_{QD}$,
given the fact that $n_{QD}$ varies by 15 $\%$ between sample 24a
and 24b (or sample 13 and 14a) at most. Sample 15 shows
conventional behavior, which, however, may be due to its proximity
to the wafer edge.

The observed transport behavior is qualitatively consistent with
the appearance of `extended' states in the 2DEG (with localization
lengths much longer than a few SAQD spacings), and reminiscent of
the situation recently analyzed in disordered but correlated 1D
systems. Izrailev and coworkers have demonstrated that even in a
1D system, potential correlations in a {\em disordered} structure
can give rise to true mobility edges, where the localization
length would diverge over a certain energy window. \cite{Izrailev}
Although the situation for 2D is not as firmly established, one
expects that the higher connectivity of electronic paths in 2D
would give rise to mobility edges even with weaker potential
correlations than in a corresponding 1D case, as numerical
evidence in a variety of systems suggests.
\cite{Dunlap-Schweitzer}

In this context, it is instructive to estimate the potential
correlations produced by the scattering centers (the charged dots)
on the 2DEG carriers. The entire SAQD charged array generates a
disordered Coulomb potential given by $V({\bf r}) = \sum_j
\frac{q_j}{|{\bf r}-{\bf R}_j|} e^{-\gamma|{\bf r}-{\bf R}_j|}$,
where each dot has effectively trapped a charge $q_i$ and the bare
dot potential is assumed screened by the 2DEG (the full potential
surely includes also local strain components).  Correspondingly,
the potential correlations are given by $\langle V(x) V(x+y)
\rangle$, where the angular brackets denote an average over the
dot population.  For a random charge distribution, $\langle q_i
q_j \rangle = \delta_{ij} q^2$, and a random distribution of dots,
the potential correlator has an asymptotic behavior given by
$\approx v_0^2 e^{-y/y_0}(a / y)$, where the correlation length
$y_0$ is a monotonic function of the screening length
$\gamma^{-1}$, which in turn is a function of electron density in
the system, and $a$ is the mean dot separation. One expects that
as the electron density increases and provides better screening,
the characteristic $y_0$ also increases, resulting in stronger
(longer range) correlations in the system. As this expression
neglects the built-in effective interaction between dots
introduced by the strain fields, as well as the correlations in
potential strength of the charged-dot landscape, this result
provides a pessimistic estimate of the true correlator in our
system. Nevertheless, even such a `weak' exponential correlator in
a 1D system can be shown to result in increasing localization
lengths for larger $y_0$ values, as shown in the lower inset of
Fig.\ 2. \cite{Izrailev} The rather strong potential correlations
that exist in our 2D system are then anticipated to yield long
localization lengths which clearly exceed the phase coherence
length of the system, and explain the complex behavior we
observe.\\
Within this interpretation, the character of the MIT observed in
our system is quite different from those reported in other
systems. One striking feature of the MIT observed in, e.g.,
Si-MOSFETs is its suppression in magnetic fields applied parallel
to the plane of the electron gas. This suppression has been
attributed to complete spin alignment.\cite{Vitkalov2000} This is
in sharp contrast to our system, where the MIT is present up to
the strongest magnetic field accessible to us, i.e. 12 T.\\
A positive magnetoresistivity around $B=0$ is also present in the
metallic regime. This resistivity minimum is robust as the
temperature is increased (Fig.\ 1b), indicating a classical
origin. Furthermore, the minimum vanishes as $n_{e}$ is reduced
into the insulating regime (Fig.\ 1a). Similar effects are
well-known from electron gases with double subband
occupancy:\cite{Houten1988} due to the different densities and
mobilities of the electrons in the two subbands, a positive
magnetoresistance evolves. Since we cannot determine the
individual electron densities in the two types of states, we
cannot fit $\rho_{xx}(B)$. Qualitatively, however, it is known
that this effect is strong if the difference in the mobilities is
significant, and consistent here with the presence of a `band' of
high mobility (extended states) next to another `band' with low
mobility.\cite{Houten1988}

In conclusion, the following qualitative picture emerges: the
presence of AAS oscillations indicates some degree of short-range
ordering, possibly in a hexagonal geometry, of the potential
landscape induced by the InAs SAQDs. This leads to the formation
of a different class of states, which may be seen as an
extended-state ``miniband'' arising from the local ordering in the
dot potential. These states become populated as $n_{e}$ is
increased above a threshold density, which corresponds
approximately to the fixed point of the MIT. For higher electron
densities, scattering between two bands with quite different
mobilities can occur, which reduces the overall mobility. The
positive magnetoresistivity around $B=0$ indicates such a
scenario. Our data thus represent the first transport signature
that two-dimensional, correlated potential landscapes can be
formed by self-ordering of charge between self-assembled quantum
dots, and raises several questions about mobility edges and
miniband design by self-ordering in two dimensions, which will
hopefully be addressed in future experiments and theoretical
studies.

It is a pleasure to thank A.K. Geim, A. Krokhin, and W. Zhang for
stimulating discussions. Financial support from ETH Z\"urich is
gratefully acknowledged. SEU acknowledges support from US DOE. ER
acknowledges support from FAPESP.

\end{multicols}

\begin{references}
\bibitem[*]{byline1}  Present address: Fakult\"at f\"ur Physik,
Universit\"at Freiburg, 79104 Freiburg, Germany.
\bibitem[\dagger]{byline2}Present address: Laborat\'{o}rio Nacional de Luz S\'{i}ncrotron, CP 6192,
    13083-061 Campinas - SP, Brazil.
 \bibitem{Leonard1993}D. Leonard, M. Krishnamurthy, C. M. Reaves, S. P.
Denbaars, and P. M. Petroff, Appl. Phys. Lett. {\bf 63}, 3203
(1993).
\bibitem{Moison1994} J. M. Moison, F. Houzay, F. Barthe, L. Leprince, E.
Andr\'{e}, and O. Vatel, Appl. Phys. Lett. {\bf 64}, 196 (1994).
\bibitem{Marzin1994} J.-Y. Marzin, J.-M. G\'{e}rard, A. Izra\"{e}l, D.
Barrier, and G. Bastard, Phys. Rev. Lett. {\bf 73}, 716 (1994).
\bibitem{Drexler1994} H. Drexler, D. Leonard, W. Hansen, J. P.
Kotthaus, and P. M. Petroff, Phys. Rev. Lett. {\bf 73}, 2252
(1994).
\bibitem{Xie1995}Q. Xie, A. Madhukar, P. Chen, and N. Kobayashi,
Phys. Rev. Lett. {\bf 75}, 2542 (1995).
\bibitem{Tersoff1996}J. Tersoff, C. Teichert, and M. Lagally,
Phys. Rev. Lett. {\bf 76}, 1675 (1996).
\bibitem{Romanov1999}A.E. Romanov, P.M. Petroff, and J.S. Speck, Appl. Phys. Lett. {\bf
74}, 2280 (1999).
\bibitem{Pinczolitis1999}M. Pinczolitis, G. Springholz, and G. Bauer,
Phys. Rev. B {\bf 60}, 11524 (1999).
\bibitem{Lee2000}H. Lee, J.A. Johnson, J.S. Speck, and P.M. Petroff,
J. Vac. Sci. Technol. B {\bf 18}, 2193 (2000).
\bibitem{Springholz1998}G. Springholz, V. Holy, M. Pinczolitis, and
G. Bauer, Science {\bf 282}, 734 (1998).
\bibitem{Ribeiro1998} E. Ribeiro, E. M\"{u}ller, T. Heinzel, H. Auderset,
K. Ensslin, G. Medeiros-Ribeiro, and P. M. Petroff, Phys. Rev. B
{\bf 58}, 1506 (1998).
\bibitem{Ribeiro1999} E. Ribeiro, R. D. J\"{a}ggi, T. Heinzel,
K. Ensslin, G. Medeiros-Ribeiro, and P. M. Petroff, Phys. Rev.
Lett. {\bf 82}, 996 (1999).
\bibitem{Buks1994}E. Buks, M. Heiblum, and H. Shtrikman, Phys. Rev. B
{\bf 49}, 14790 (1994).
\bibitem{Umbach1986}C.P. Umbach, C.v. Haesendonck, R.B. Laibowitz,
S. Washburn, and R.A. Webb, Phys. Rev. Lett. {\bf 56}, 386 (1986).
\bibitem{Spivak1981} B.L. Altshuler, A.G. Aronov, and B.Z. Spivak, JETP
Lett. {\bf 33}, 94  (1981).
\bibitem{Nihey1995}F. Nihey, S. W. Hwang, and K. Nakamura, Phys. Rev. B
 {\bf 51}, 4649 (1995).
\bibitem{Heinzel2000} This estimation of $n_{QD}$ from transport experiments assumes
that two electrons are captured per dot. Higher occupation numbers
are possible and reduce the value extracted for $n_{QD}$.
\bibitem{Lee2001}H. Lee, J.A. Johnson, M.Y. He, J.S. Speck,and P.M. Petroff,
Appl. Phys. Lett. {\bf 78}, 105 (2001).
\bibitem{Izrailev} F.M. Izrailev and A.A. Krokhin, \prl {\bf 82},
4062 (1999); F.M. Izrailev {\em et al}., \prb {\bf 63}, 041102(R)
(2001).
\bibitem{Dunlap-Schweitzer} See, for example, D.H. Dunlap, K. Kundu
and P. Phillips, \prb {\bf 40}, 10999 (1989); Th. Koschny, H.
Potempa, and L. Schweitzer, \prl {\bf 86}, 3863 (2001).
\bibitem{Vitkalov2000}S.A. Vitkalov, H. Zheng, K.M. Mertes, M.P.
Sarachik, and T.M. Klapwijk,Phys. Rev. Lett. {\bf 85}, 2164
(2000).
\bibitem{Houten1988}H. van Houten, J.G. Williamson, M.E.I.
Broekaart, C.T. Foxon, and J.J. Harris, Phys. Rev. B {\bf 37},
2756 (1988).
\end{references}
\end{document}